\documentclass[12pt,preprint]{aastex}
\RequirePackage{lineno}
\usepackage{amssymb}
\usepackage{amsmath}
\usepackage{mathrsfs}
\usepackage{natbib}
\usepackage[colorlinks, linkcolor=blue, urlcolor=blue, citecolor=blue]{hyperref}
\usepackage{array}
\usepackage{float}
\usepackage{graphicx}
\usepackage{subfigure}
\usepackage{rotating}
\usepackage{color}
\usepackage[all]{hypcap}
\usepackage[toc,page]{appendix}

\begin{document}

\title{A Multiple Ejecta-Circumstellar Medium Interaction Model and Its
Implications for Superluminous Supernovae iPTF15esb and iPTF13dcc}
\author{Liang-Duan Liu\altaffilmark{1,2,3}, Ling-Jun Wang\altaffilmark{4},
Shan-Qin Wang\altaffilmark{1,2,5}, and Zi-Gao Dai\altaffilmark{1,2}}

\begin{abstract}

In this paper, we investigate two hydrogen-poor superluminous
supernovae (SLSNe) iPTF15esb and iPTF13dcc whose light curves (LCs) show
significant deviation from the smooth rise and fall. The LC of iPTF15esb exhibits
two peaks and a post-peak plateau, and furthermore the late-time spectrum of
iPTF15esb shows a strong, broad H$\alpha$ emission line. The early-time LC of
iPTF13dcc shows a long duration bump followed by the second peak.
Here we propose an ejecta-circumstellar medium (CSM) interaction model
involving multiple shells/winds and use it to explain the LCs of iPTF15esb
and iPTF13dcc. We find that the theoretical LCs reproduced by this model
can well match the observations of iPTF15esb and iPTF13dcc. Based on our
results, we infer that the progenitors have undergone multiple violent
mass-loss processes before the SN explosion.
In addition, we find that the variation trend of our inferred densities
of the shells is consistent with that predicted by the stellar mass-loss history
before an SN explosion. Further investigations for other bumpy SLSNe/SNe
would shed light on their nature and provide a probe for the mass-loss
history of their progenitors.
\end{abstract}


\affil{\altaffilmark{1}School of Astronomy and Space Science,
Nanjing University, Nanjing 210093, China; dzg@nju.edu.cn}
\affil{\altaffilmark{2}Key Laboratory of Modern Astronomy and
Astrophysics (Nanjing University), Ministry of Education, China}
\affil{\altaffilmark{3}Department of Physics and Astronomy,
University of Nevada, Las Vegas, NV 89154, USA}
\affil{\altaffilmark{4}Astroparticle Physics,
Institute of High Energy Physics,
Chinese Academy of Sciences, Beijing 100049, China}
\affil{\altaffilmark{5}Department
of Astronomy, University of California, Berkeley, CA 94720-3411,
USA}

\section{Introduction}

\label{sec:Intro}

In the past decade, fast-developing non-targeted supernova (SN) survey
programs have discovered a new class of unusual SNe whose peak absolute
magnitudes $M_{\mathrm{peak}}$ are brighter than $-21$ mag.
These very luminous SNe are called ``superluminous supernovae (SLSNe)" %
\citep{Qiu2011,Gal2012}.

It appears that SLSNe can be simply divided into two subclasses, types I and
II. SLSNe I have spectra around the peaks that are lack of hydrogen
absorption lines and their light curves (LCs) might be explained by the pair
instability SNe \citep[PISNe;][]{Rak1967,Heg2002,Heg2003}, the $^{56}$Ni-powered
model, the magnetar-powered model
\citep{Kas2010,Woo2010,Ins2013,Wang2015a,Wang2015b,Dai2016,Wang2016a,Liu2017,Yu2017},
or the ejecta-circumstellar medium (CSM) interaction model %
\citep{Che2011,Cha2012,Cha2013,Gin2012,Nic2014,Chen2015}.

On the other hand, the spectra around the peaks of SLSNe II show strong
hydrogen emission features and almost all of them show narrow and
intermediate width Balmer emission lines, similar to normal SNe IIn.
Previous studies \citep{Sim2007,Mor2011,Cha2012,Cha2013,Mor2013} suggested
that the LCs of SLSNe IIn might be powered by the interactions between the
SN ejecta and the dense, hydrogen-rich, and optically thick CSM.

However, some SLSNe I (e.g., iPTF13ehe, iPTF15esb and iPTF16bad) whose
late-time spectra exhibit H$\alpha$ emission lines \citep{Yan2015,Yan2017}
complicated the classification scheme. \citet{Yan2015} estimated that $15\%$
of SLSNe I might have these spectral features. Among these SLSNe I that have
late-time H$\alpha$ emission lines, iPTF15esb, exploded at a redshift $z$ of
0.224, is the most striking one. Its late-time spectra show strong,
broad H$\alpha $ emission lines, indicating the interaction between the
SN ejecta and the hydrogen-rich CSM shell surrounding the SN progenitor.
Moreover, its LC has two peaks whose luminosities are approximately equal to
each other ($L_{\text{peaks}}\approx 4\times 10^{43}$ erg s$^{-1}$) and a
plateau lasting about 40 days. Its late-time LC decays as $L_{\text{bol}
}\propto t^{-2.5}$. Another interesting case is iPTF13dcc \citep{Vre2017}
whose LC shows an initial slow decline with a duration being $\sim 30$\,days, and
then rebrightens and reaches its second peak.

Both the LCs of iPTF15esb and iPTF13dcc challenge all the models mentioned above.
The decline rate of the late-time LCs powered by $^{56}$Ni cascade decay with
full trapping of $\gamma$ rays is 0.0098 mag per day and the late-time LCs
powered by a magnetar (with full trapping of $\gamma $-rays) can be
described by $L_{\text{inp,mag}}\propto t^{-2}$. The magnetar model together
with $^{56}$Ni cascade decay with leakage of $\gamma$ rays %
\citep{Clo1997,Wang2015a,Chen2015} is able to explain the late-time LC of
iPTF15esb, but this model cannot yield the LC showing two bright peaks and a plateau.
It seems that an energy-source model involving multiple energy injections is
needed to account for the exotic LCs of iPTF15esb.
\cite{Wang2016b} proposed a triple-energy-source model ($^{56}$Ni plus
magnetar plus interaction) and used it to explain the LC of iPTF13ehe.
However, this model involves only one collision between the ejecta and the
CSM shell and cannot produce the undulatory LCs.
The model containing cooling emission and magnetar spinning-down or
the ejecta-CSM interaction that has been adopted by \citet{Vre2017} to
fit the double-peaked LC of iPTF13dcc cannot yet reproduce
such undulations seen in the LC of iPTF15esb.

As pointed out by \citet{Yan2017}, however, the spectrum and the LC seem to
favor the interactions between the SN ejecta and multiple
CSM shells or CSM clumps at different radii.
Here we propose an ejecta-CSM interaction model involving interactions
between the SN ejecta and multiple shells and stellar winds and use this
model to fit the LC of iPTF15esb. The double interaction model is also
promising to account for the LC of iPTF13dcc.

This paper is structured as follows. In Section \ref%
{sec:mod}, we give a detailed description of the model, and apply it to fit
the LCs of iPTF15esb and iPTF13dcc in Section \ref{sec:res}. Finally, we
discuss our results and present our conclusions in Section \ref{sec:dis}.

\section{Multiple Ejecta-CSM Interaction Model}

\label{sec:mod}

In this section, we generalize the normal ejecta-CSM interaction model to more
complicated model that involves multiple CSM shells and winds. The basic
physical picture of this model is described below.
The interaction of the ejecta with the pre-existing CSM results in the
formation of two shock waves: a forward shock (FS) propagating through the
CSM and a reverse shock (RS) sweeping up the SN ejecta %
\citep{Che1982,Che1994}. The interaction provides a strong energy source by
converting the kinetic energy to radiation.

Based on numerical simulations for an SN explosion, a broken power-law
distribution for the density of the SN ejecta can be adopted \citep{Mat1999}%
. The density profile of the outer part ejecta is
\begin{equation}
\rho _{\text{NS,out}}=g_{n}t^{n-3}r^{-n},
\end{equation}%
where $n$ is the slope of the outer part ejecta, depending on the SN progenitor
star, and $g_{n}$ is the density profile scaling parameter, which is given
by \citep{Che1994,Cha2012}
\begin{equation}
g_{n}=\frac{1}{4\pi \left( n-\delta \right) }\frac{\left[ 2\left( 5-\delta
\right) \left( n-5\right) E_{\text{SN}}\right] ^{\left( n-3\right) /2}}{%
\left[ \left( 3-\delta \right) \left( n-3\right) M_{\text{ej}}\right]
^{\left( n-5\right) /2}},
\end{equation}
where $\delta $ is the inner density profile slope. Here $E_{\text{SN}}$ is
the total SN energy, and $M_{\text{ej}}$ is the total mass of the SN ejecta.
The relation between $E_{\text{SN}}$ and $M_{\text{ej}}$ can be written as %
\citep{Cha2012}
\begin{equation}
E_{\text{SN}}=\frac{(3-\delta)\left( n-3\right) }{2\left( 5-\delta \right) \left(
n-5\right) }M_{\text{ej}}\left( x_{0}v_{\text{SN}}\right) ^{2},
\end{equation}
where $x_0$ denotes the dimensionless radius of a break in the supernova
ejecta density profile from the inner component to the outer component.

Before the SN explosion, the mass loss of a massive star could erupt several
gas shells surrounding the progenitor. We assume that the density of a
circumstellar shell or wind is
\begin{equation}
\rho _{\text{CSM,i}}=q_{\text{i}}r^{-s_{\text{i}}},
\end{equation}%
where $q_{\rm i}$ is a scaling constant, and $s_{\rm i}$ is the power-law index for CSM
density profile and therefore $s_{\rm i}=2$ indicates stellar winds while $s_{\rm i}=0$
indicates uniform density shells. The subscript ``i'' denotes the $i$th
collision between the ejecta and the CSM shell. For a steady wind ($s_{\rm i}=2$)
with a constant pre-explosion mass loss rate $\dot{M}$ and wind velocity $v_{%
\text{w}}$, we have $q=\dot{M}/\left( 4\pi v_{\text{w}}\right) $.

The shocked CSM and shocked ejecta are separated by a contact discontinuity.
The radius of the contact discontinuity $R_{\text{cd}}$ can be described by a
self-similar solution \citep{Che1982}
\begin{equation}
R_{\text{cd,i}}=\left( \frac{A_{\text{i}}g_{n}}{q_{\text{i}}}\right) ^{\frac{%
1}{n-s_{\text{i}}}}t^{\frac{\left( n-3\right) }{\left( n-s_{\text{i}}\right)
}},
\end{equation}%
where $A_{\rm i}$ is a constant. The radii of the FS and RS are given by
\begin{equation}
R_{\text{FS,i}}\left( t\right) =R_{\text{in,i}}+\beta _{\text{FS,i}}R_{\text{%
cd,i}}
\end{equation}%
and
\begin{equation}
R_{\text{RS},\text{i}}\left( t\right) =R_{\text{in,i}}+\beta _{\text{RS,i}%
}R_{\text{cd,i}},
\end{equation}%
where $R_{\text{in,i}}$ is the initial radius of the $i$th interaction (which is equal to the inner radius
of the CSM density profile), $\beta _{\text{FS}}$ and $\beta _{\text{RS}}$ are
constants representing the ratio of the shock radii to the
contact-discontinuity radius $R_{\text{cd}}$. The values of $\beta _{\text{FS%
}}$ and $\beta _{\text{RS}}$ are determined by the values of $n$ and $s_{\rm i}$.
They are given in Table 1 of \cite{Che1982}. For $n=7$ and $s_{\rm i} = 2$, we can
obtain $\beta _{\text{FS}} = 1.299$, $\beta _{\text{RS}}=0.970$, and $A = 0.27$;
for $n=7$ and $s_{\rm i} = 0$, we have $\beta _{\text{FS}}=1.181$, $\beta_{\text{RS}%
}=0.935$, and $A=1.2$.

The interaction radii which are equal to the inner radii of the CSM density profile
are given by
\begin{equation}
R_{\text{in,i}}=R_{\text{in,i-1}}+\left( t_{\text{tr,i}}-t_{\text{tr,i-1}%
}\right) \left( \frac{2\left( 5-\delta \right) \left( n-5\right) E_{\text{k,i%
}}}{x_{0}^{2}(3-\delta)\left( n-3\right) M_{\text{ej,i}}}\right) ^{1/2},
\label{eq: Rin}
\end{equation}%
where $t_{\text{tr,i}}$ is the trigger time of the $i$th interaction relative to time
zero point. Here, we set the first interaction between ejecta and CSM as the time zero point.
The kinetic energy of the $i$th interaction is
\begin{equation}
E_{\text{k,i}}=E_{\text{k,i}-1}-E_{\text{rad,i}-1},
\end{equation}%
where $E_{\text{rad}}$ is the energy loss due to radiation. The ejecta mass
of the $i$th interaction is
\begin{equation}
M_{\text{ej,i}}=M_{\text{ej,i}-1}+M_{\text{CSM,i}-1}  \label{eq: Mej}.
\end{equation}

The interaction between the ejecta and the CSM would convert the kinetic
energy to radiation. The luminosity input function of the forward shock is \citep{Cha2012}
\begin{equation}
L_{\text{FS,i}}\left( t\right) =\frac{2\pi }{\left( n-s_{\text{i}}\right)
^{3}}g_{n}^{\frac{5-s_{\text{i}}}{n-s_{\text{i}}}}q_{\text{i}}^{\frac{n-5}{%
n-s_{\text{i}}}}\left( n-3\right) ^{2}\left( n-5\right) \beta _{\text{FS,i}%
}^{5-s_{\text{i}}}A_{\text{i}}^{\frac{n-5}{n-s_{\text{i}}}}\left( t+t_{\text{%
int,i}}\right) ^{\alpha _{\text{i}}}\theta \left( t_{\text{FS},\text{BO,i}%
}-t\right),
\end{equation}%
while the reverse shock's input luminosity is \citep{Wang2017}\footnote{%
For reverse shock luminosity we use the expression given by \cite{Wang2017},
instead of that given by \cite{Cha2012}. See \cite{Wang2017} for more details.}
\begin{equation}
L_{\text{RS,i}}\left( t\right) =2\pi \left( \frac{A_{\text{i}}g_{n}}{q_{%
\text{i}}}\right) ^{\frac{5-n}{n-s_{\text{i}}}}\beta _{\text{RS,i}%
}^{5-n}g_{n}\left( \frac{n-5}{n-3}\right) \left( \frac{3-s_{\text{i}}}{n-s_{%
\text{i}}}\right) ^{3}\left( t+t_{\text{int,i}}\right) ^{\alpha _{\text{i}%
}}\theta \left( t_{\text{RS},\ast \text{,i}}-t\right) ,
\end{equation}%
where $\theta \left( t_{\text{RS},\ast }-t\right) $ and $\theta \left( t_{%
\text{FS},\text{BO}}-t\right) $ represent the Heaviside step function that
control the end times of FS and RS, respectively, and $t_{\text{int,i}}\approx R_{%
\text{in,i}}/v_{\text{SN,i}}$ is the time when the ejecta-CSM interaction
begins. The temporal index is $\alpha_i =\left( 2n+6s_{\rm i}-ns_{\rm i}-15\right) /\left(
n-s_{\rm i}\right) $. Here we fix $n=7$. Consequently, we have $\alpha_i =-0.143$ for
the shell ($s_{\rm i}=0$), and $\alpha_i =-0.6$ for the steady wind ($s_{\rm i}=2$).

The RS termination timescale $t_{%
\text{RS},\ast }$ is the time once the RS sweeps up all available ejecta \citep{Cha2012,Cha2013}
\begin{equation}
t_{\text{RS,}\ast ,\text{i}}=\left[ \frac{v_{\text{SN,i}}}{\beta _{\text{RS,i%
}}\left( A_{\text{i}}g_{n}/q_{\text{i}}\right) ^{\frac{1}{n-s_{\text{i}}}}}%
\left( 1-\frac{\left( 3-n\right) M_{\text{ej,i}}}{4\pi v_{\text{SN,i}%
}^{3-n}g_{n}}\right) ^{\frac{1}{3-n}}\right] ^{\frac{n-s_{\text{i}}}{s_{%
\text{i}}-3}}.
\end{equation}

Under the same assumption, the FS terminates when the optically thick part
of the CSM is swept up. The termination timescale of the FS, being
approximately equal to the time of FS breakout $t_{\text{FS},\text{BO}}$, is
given by \citep{Cha2012,Cha2013}
\begin{equation}
t_{\text{FS,BO,i}}=\left\{ \frac{\left( 3-s_{\text{i}}\right) q_{\text{i}%
}^{\left( 3-n\right) /\left( n-s_{\text{i}}\right) }\left[ A_{\text{i}}g_{n}%
\right] ^{\left( s_{\text{i}}-3\right) /\left( n-s_{\text{i}}\right) }}{4\pi
\beta _{\text{FS,i}}^{3-s_{\text{i}}}}\right\} ^{\frac{n-s_{\text{i}}}{%
\left( n-3\right) \left( 3-s_{\text{i}}\right) }}M_{\text{CSM,th,i}}^{\frac{%
n-s_{\text{i}}}{\left( n-3\right) \left( 3-s_{\text{i}}\right) }},
\end{equation}%
where $M_{\text{CSM,th,i}}$ is the mass of optically thick CSM
\begin{equation}
M_{\text{CSM,th,i}}=\int_{R_{\text{in,i}}}^{R_{\text{ph,i}}}4\pi r^{2}\rho _{%
\text{CSM,i}}dr.
\end{equation}%
Here $R_{\text{ph,i}}$ denotes the photospheric radius of the $i$th CSM
shell, located at the optical depth $\tau =2/3$ under Eddington's
approximation. $R_{\text{ph,i}}$ is given by
\begin{equation}
\tau =\int_{R_{\text{ph,i}}}^{R_{\text{out,i}}}\kappa _{\text{i}}\rho _{%
\text{CSM,i}}dr=\frac{2}{3},
\end{equation}%
where $\kappa $ is the optical opacity of the CSM and $R_{\text{out,i}}$ is
the radius of the outer boundary of the CSM. $R_{\text{out,i}}$ can be determined by
\begin{equation}
M_{\text{CSM,i}}=\int_{R_{\text{in,i}}}^{R_{\text{out,i}}}4\pi r^{2}\rho _{%
\text{CSM,i}}dr.
\end{equation}

Both the FS and the RS heat the interacting material. The total luminosity
input from the FS and RS can be written as
\begin{equation}
L_{\text{inp,CSM,i}}\left( t\right) =\epsilon _{\text{i}}\left[ L_{\text{FS,i%
}}\left( t\right) +L_{\text{RS,i}}\left( t\right) \right] ,
\end{equation}%
where $\epsilon _{\text{i}} $ is the conversion efficiency from the kinetic energy. \cite%
{Cha2012} assumed that $\epsilon =$ 100$\%$, which is unrealistic in the
actual situation, especially in the $M_{\text{CSM}}\ll M_{\text{ej}}$ case.
Due to the poor knowledge of the process of converting the kinetic energy to
radiation, for simplicity, we set $\epsilon _{\text{i}} $ as a free parameter.

Because the expansion velocity of the CSM is much lower than the typical
velocity of the SN ejecta, \cite{Cha2012} assumed a fixed photosphere inside the
CSM. Under this assumption, the output bolometric LC can be written as
\begin{equation}
L_{\text{i}}\left( t\right) =\frac{1}{t_{\text{diff,i}}}\exp \left[ -\frac{t%
}{t_{\text{diff,i}}}\right] \int_{0}^{t}\exp \left[ \frac{t^{\prime }}{t_{%
\text{diff,i}}}\right] L_{\text{inp,CSM,i}}\left( t^{\prime }\right)
dt^{\prime },
\end{equation}%
where $t_{\text{diff,i}}$ is the diffusion timescale in the optically thick
CSM. The diffusion timescales of the $i$th interaction can be written as
\begin{equation}
t_{\text{diff,i}}=\sum_{j=i}^{N}\frac{\kappa _{\text{j}}M_{\text{CSM,th,j}}}{%
\beta cR_{\text{ph}}}.
\end{equation}%
where $\beta =4\pi ^{3}/9\simeq 13.8$ is a constant \citep{Arn1982}, and $c$
is the speed of light.

In this multiple interaction model, the theoretical bolometric LC of
$N$ times interactions can be described as
\begin{equation}
L_{\text{tot}}\left( t\right) =\sum_{i=1}^{N}L_{\text{i}}\left(t\right) .
\end{equation}

We assume that the bolometric luminosity comes from the blackbody emission
from the photosphere whose radius is $R_{\text{ph}}$, and therefore the
temperature in our model can be estimated by
\begin{equation}
T=\left( \frac{L_{\text{tot}}}{4\pi R_{\text{ph}}^{2}\sigma_{\rm SB}}\right) ^{1/4},
\end{equation}%
where $\sigma_{\rm SB}$ is the Stefan-Boltzmann constant. By assuming a stationary
photosphere, we have $T\propto L_{\text{tot}}^{1/4}$.

\section{Modeling the light curves of SLSNe with multiple peaks}

\label{sec:res}

In this section, we use the model described above to fit the bolometric
light curves of iPTF15esb and iPTF13dcc.
In order to derive the best-fitting parameters and determine the
ranges of relevant parameters, we develop a Markov Chain Monte Carlo (MCMC)
method that can minimize the values of $\chi^{2}$ divided by
the number of degree of freedom ($\chi^2$/dof) for the multiple ejecta-CSM
interaction model and employ this model to fit the LCs of iPTF15esb and iPTF13dcc.

To reduce the number of free parameters in our calculations,
we fix several parameters. We adopt the power-law index of the outer
density profile $n=7$ as an approximation for Type I SNe \citep{Che1982}, and
the inner density slope $\delta =0$.

The free parameters in our model are the opacities of the CSM
shells and winds $\kappa$, the mass of the SN ejecta $M_{\text{ej}}$, the
total mass of the CSM $M_{\text{CSM}}$, the density of the CSM at the
interaction radius $\rho _{\text{CSM,in}}$, the interaction radius (the
inner radius of CSM) $R_{\text{in}}$, the conversion efficiency from the
kinetic energy to radiation $\epsilon $, the time of the collision between
the SN ejecta and the CSM shells $t_{\text{tr}}$.

The opacities of the CSM shells and winds $\kappa$ are related to their
composition and temperatures. For hydrogen-poor matter, the
dominant source of opacity is electron scattering, $\kappa =0.06-0.2$ cm$%
^{2} $ g$^{-1}$ (see the references listed in \citealt{Wang2015b}). For
hydrogen-rich matter, $\kappa =0.33$ cm$^{2}$ g$^{-1}$, which is the Thomson
electron scattering opacity for fully ionized material with the solar
metallicity \citep{Mor2011,Cha2012}.

\subsection{iPTF15esb}

It is reasonable to assume that there are at least three collisions between
the SN ejecta and the CSM shells since the LC of iPTF15esb shows two prominent
peaks and a plateau. In this scenario, the interaction between the SN ejecta
and the stellar wind (i.e., $s_{1}=2$) powers the first peak of the LC of
iPTF15esb while the second peak and the plateau are powered by the interactions
between the SN ejecta and CSM shells ($s_{2}=s_{3}=0$) at different radii.

The composition of the first and second CSM shells cannot be well constrained
since no hydrogen emission lines in the early-time spectra of iPTF15esb have
been detected. On the other hand, the strong, broad H$\alpha$ emission at $\sim 70$
days after the first LC peak might be prompted by the interaction between the ejecta
and a hydrogen-rich shell.
Therefore, the values of the opacity of the first and second CSM shells $\kappa_{1}$
and $\kappa_{2}$ can be fixed to be $0.06-0.2$ cm$^{2}$ g$^{-1}$ or 0.33 cm$^{2}$ g$^{-1}$,
and the value of the opacity of the third CSM shell $\kappa_{3}$ is supposed to
be $0.33$ cm$^{2}$ g$^{-1}$. By analyzing Fe\textsc{ii} 5169 \AA\ line, \cite{Yan2017}
found that the photospheric velocity around the first peak of iPTF15esb is
$v_{\text{ph}}\approx 17,800$ km s$^{-1}$ which can be set to be the value of the
characteristic velocity $v_{\text{SN}}$ of the SN ejecta.

\begin{figure}[tbph]
\begin{center}
\includegraphics[width=0.48\textwidth,angle=0]{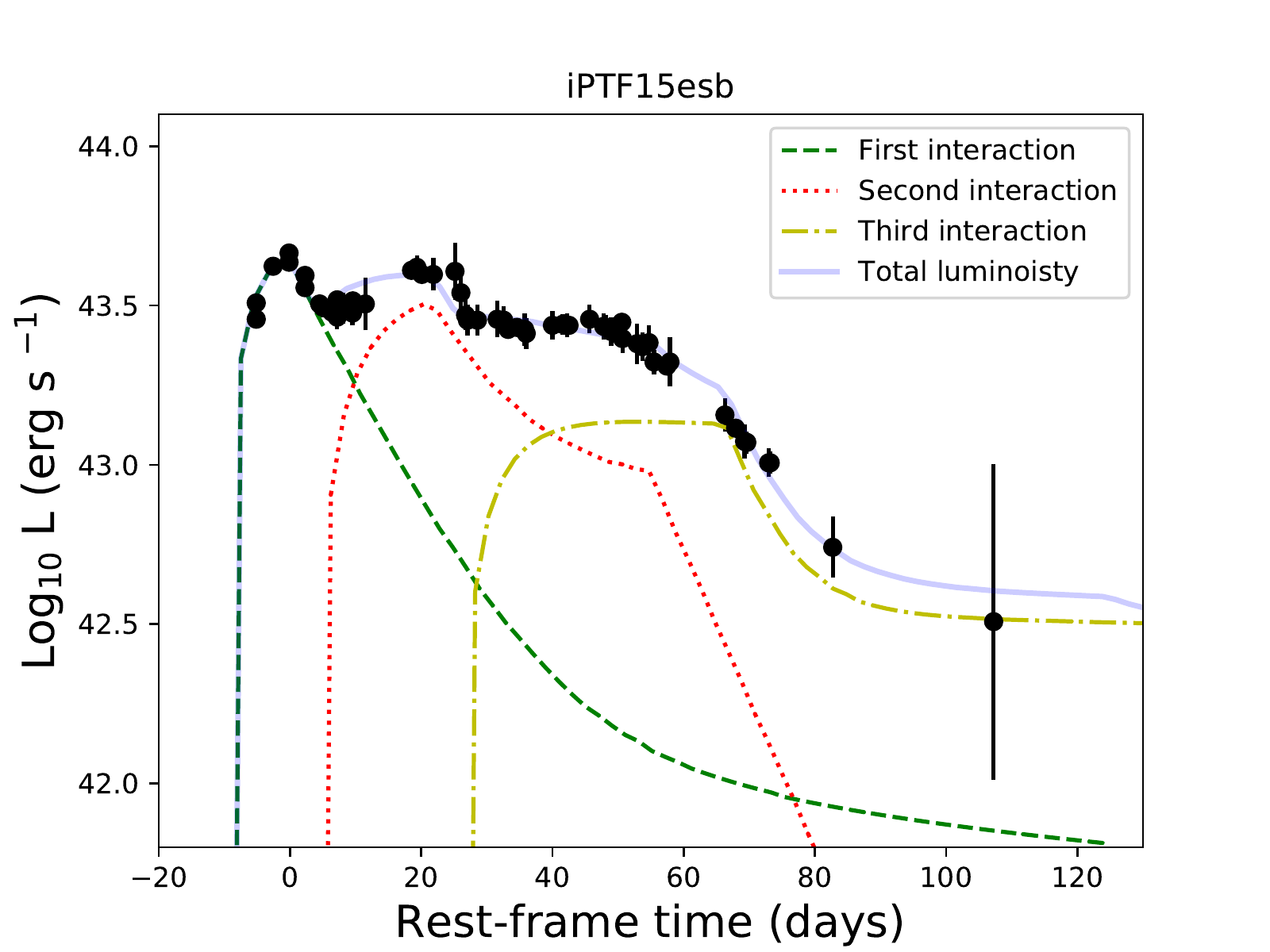}
\includegraphics[width=0.48\textwidth,angle=0]{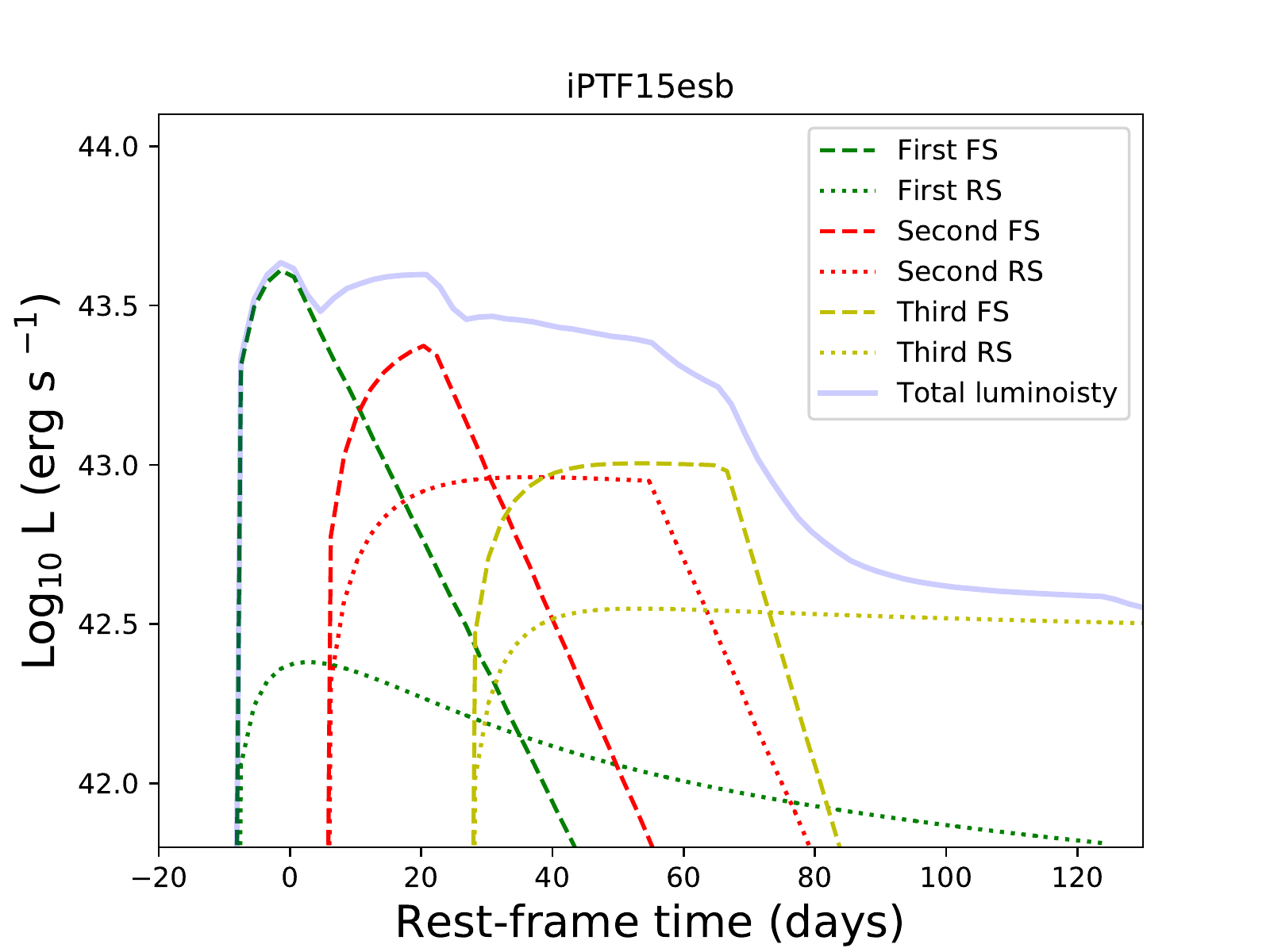}
\end{center}
\caption{Left panel: The fit to the bolometric LC of iPTF15esb using the multiple
ejecta-CSM interaction model. Data are obtained from \protect\cite{Yan2017}.
Right panel: Separate contributions of the forward shocks and
reverse shocks to the theoretical bolometric LC of iPTF15esb for
successive interactions.
The fitting parameters are shown in the text and Table \protect\ref%
{tbl:fitting par}.}
\label{fig:LC_15esb}
\end{figure}

\begin{figure}[tbph]
\begin{center}
\includegraphics[width=1.1\textwidth,angle=0]{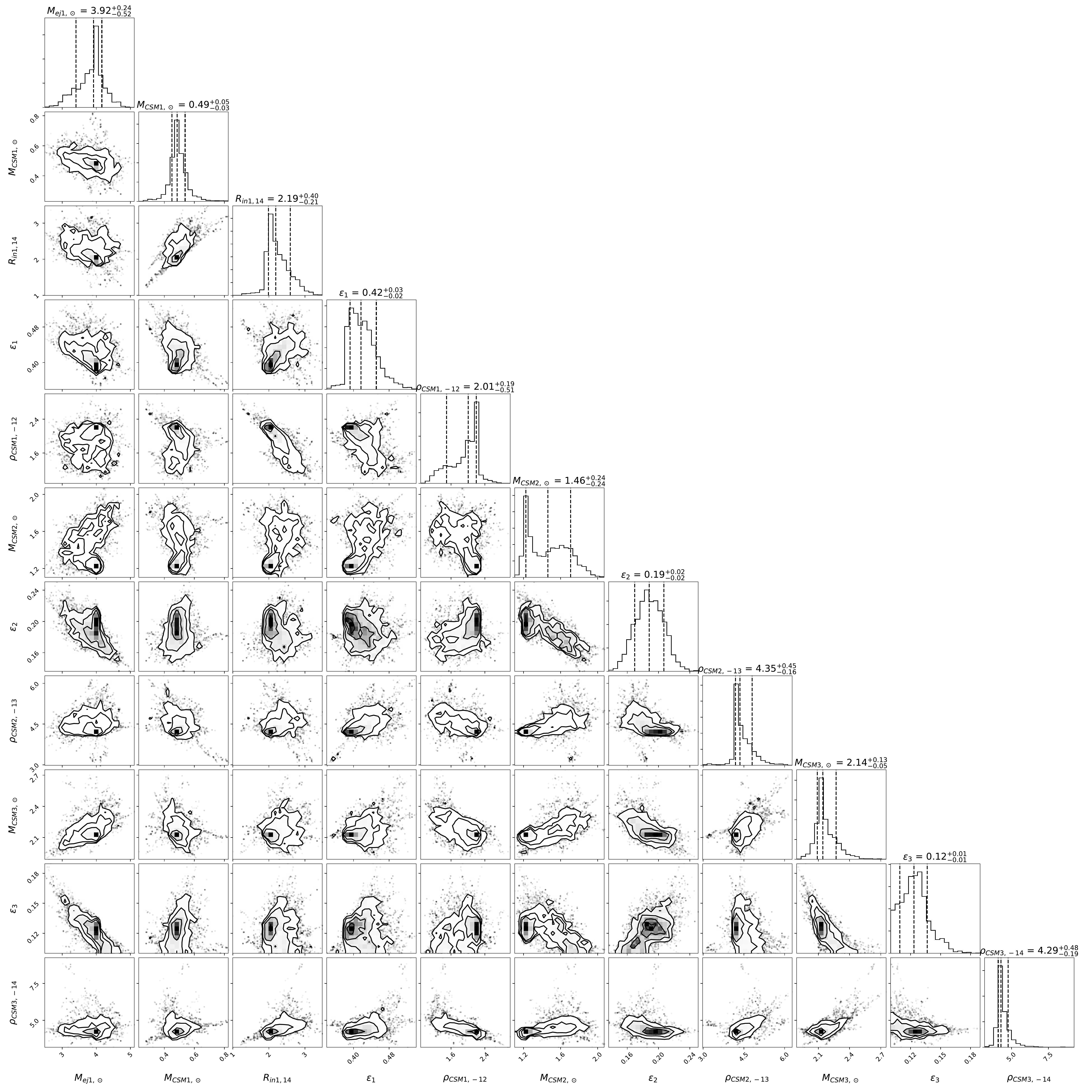}
\end{center}
\caption{Corner plot of the parameters for fitting the LC of iPTF15esb.
Medians and $1 \sigma$ ranges are labeled.}
\label{fig:corner-15esb}
\end{figure}

The theoretical LC of iPTF15esb is shown in Figure \ref{fig:LC_15esb}. 
The best-fitting parameters and the corresponding confidence
contour corners of iPTF15esb are shown in Table \ref{tbl:fitting par}
and Figure \ref{fig:corner-15esb}, respectively.
We find that the multiple interaction model can explain the bumpy LC of iPTF15esb
well ($\chi^{2}$/dof = 1.71) and the parameters are reasonable.
The derived physical parameters of the CSM shells and the wind are listed in Table %
\ref{tbl:der}. The masses of the optically thick part of the CSM shells $M_{%
\text{CSM,th}}$, which are close to their total mass, can be calculated. The
termination timescales of the FS and the RS can also be determined. The
optical depth of CSM $\tau_{\text{CSM}}>1$, indicating that these shells
are opaque.

Provided that the velocity of the
shells $v_{\text{shell}}$ is $\sim$ 100 km s$^{-1}$ and using $t_{\text{shell}%
}\approx R_{\text{in}}/v_{\text{shell}}$, we can obtain the time when the
progenitor expelled the CSM shells before explosion.
We infer that the progenitor of iPTF15esb has undergone at least two
violent shell eruption processes at 6.69, and 16.34
years before the SN explosion, respectively, then
experienced a wind-like mass loss whose mass-loss rate ($\dot{M}$)
is $0.19-1.9 M_{\odot}$ yr$^{-1}$. \footnote{$\dot{M} = 4{\pi}r^2\rho(r)v_{\text{w}}$,
where $\rho(r)$ is the average density at a radius $r$, and
$v_{\text{w}}$ $\approx 100-1,000$ km s$^{-1}$ is the terminal velocity of
stellar wind of a hydrogen-poor star (He star; \citealt{Smi2014}).}

\subsection{iPTF13dcc}

\citet{Vre2017} used a model combining the cooling emission from an
shock-heated extended envelope \citep{Piro2015} and energy injection from
a magnetar or an ejecta-CSM interaction to fit the LC of iPTF13dcc. While this
model is plausible, we suggest that the double-collision model is also a possible
model accounting for the LC of iPTF13dcc. In our scenario, the early-time
bump of iPTF13dcc might be powered by the first collision between the ejecta
and the CSM shell, while the second collision at larger radius powers the
late-time rebrightening of iPTF13dcc.

The early bump and the late-time rebrightening of iPTF13dcc are
powered by the interactions between the SN ejecta and CSM shells at
different radii, in which $s_{1}=s_{2}=0$ is adopted. In our fitting, we
adopt the expansion velocity $v_{\text{SN}}=10,000$ km s$^{-1}$,
which is the same as \citet{Vre2017}. We assume the opacities of the first
and second CSM shells $\kappa_{1}=\kappa_{2}=0.2$ cm$^{2}$ g$^{-1}$.

\begin{figure}[tbph]
\begin{center}
\includegraphics[width=0.48\textwidth,angle=0]{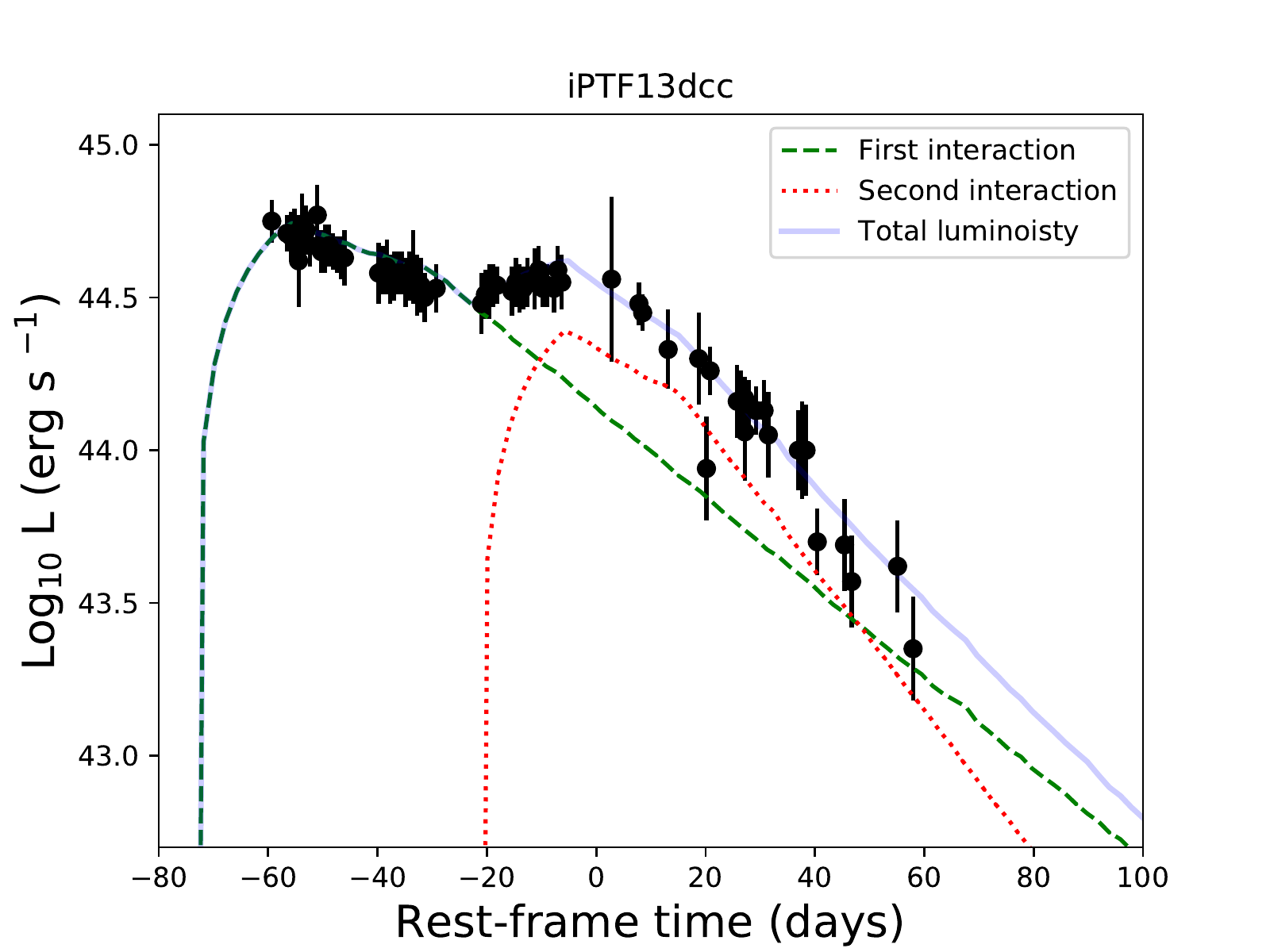}
\includegraphics[width=0.48\textwidth,angle=0]{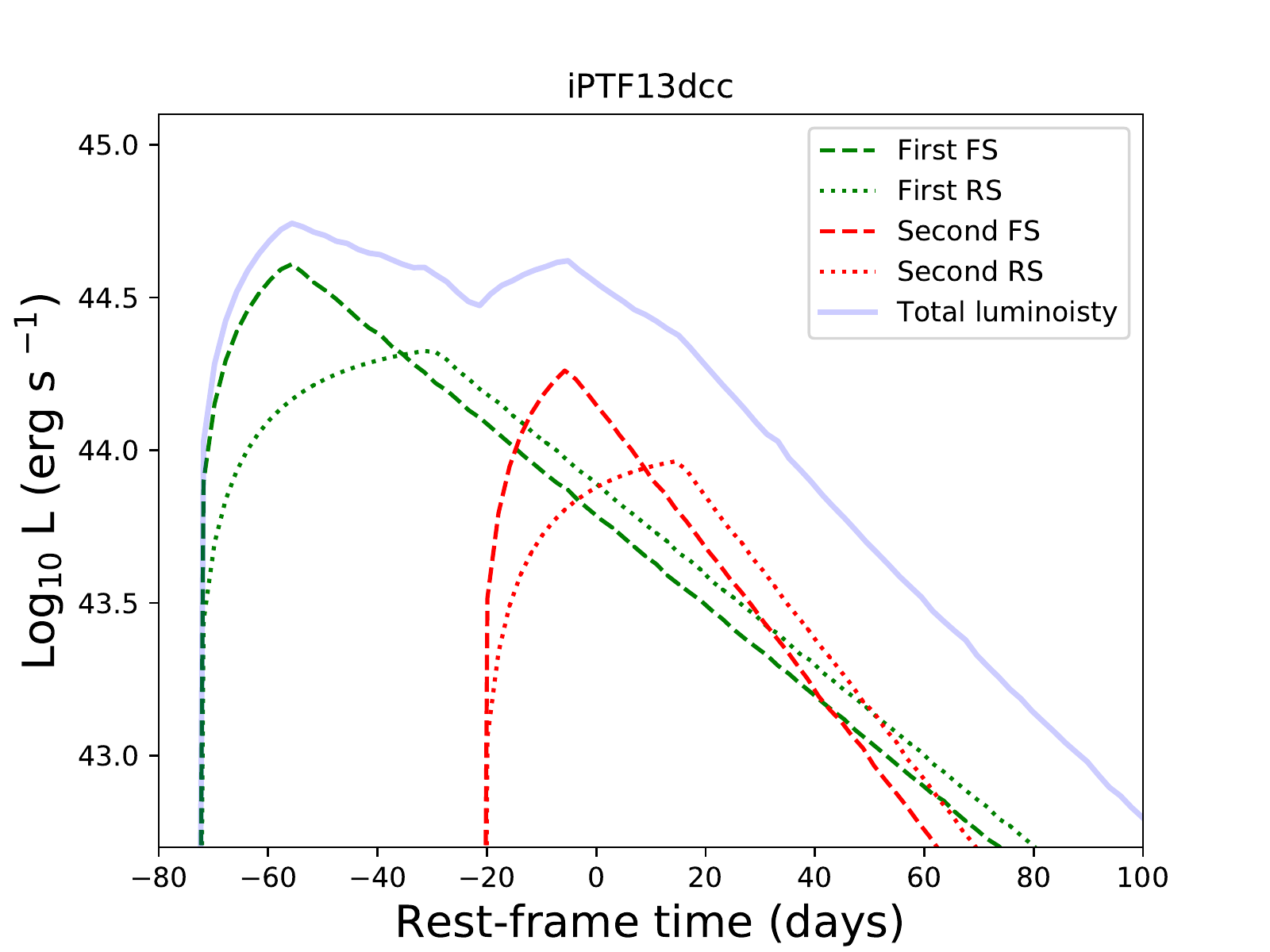}
\end{center}
\caption{Same as in Figure \protect\ref{fig:LC_15esb} but for
iPTF13dcc. Data are obtained from \protect\cite{Vre2017}.
The fitting parameters are shown in the text and Table \protect\ref%
{tbl:fitting par}.}
\label{fig:LC_13dcc}
\end{figure}

\begin{figure}[tbph]
\begin{center}
\includegraphics[width=1.1\textwidth,angle=0]{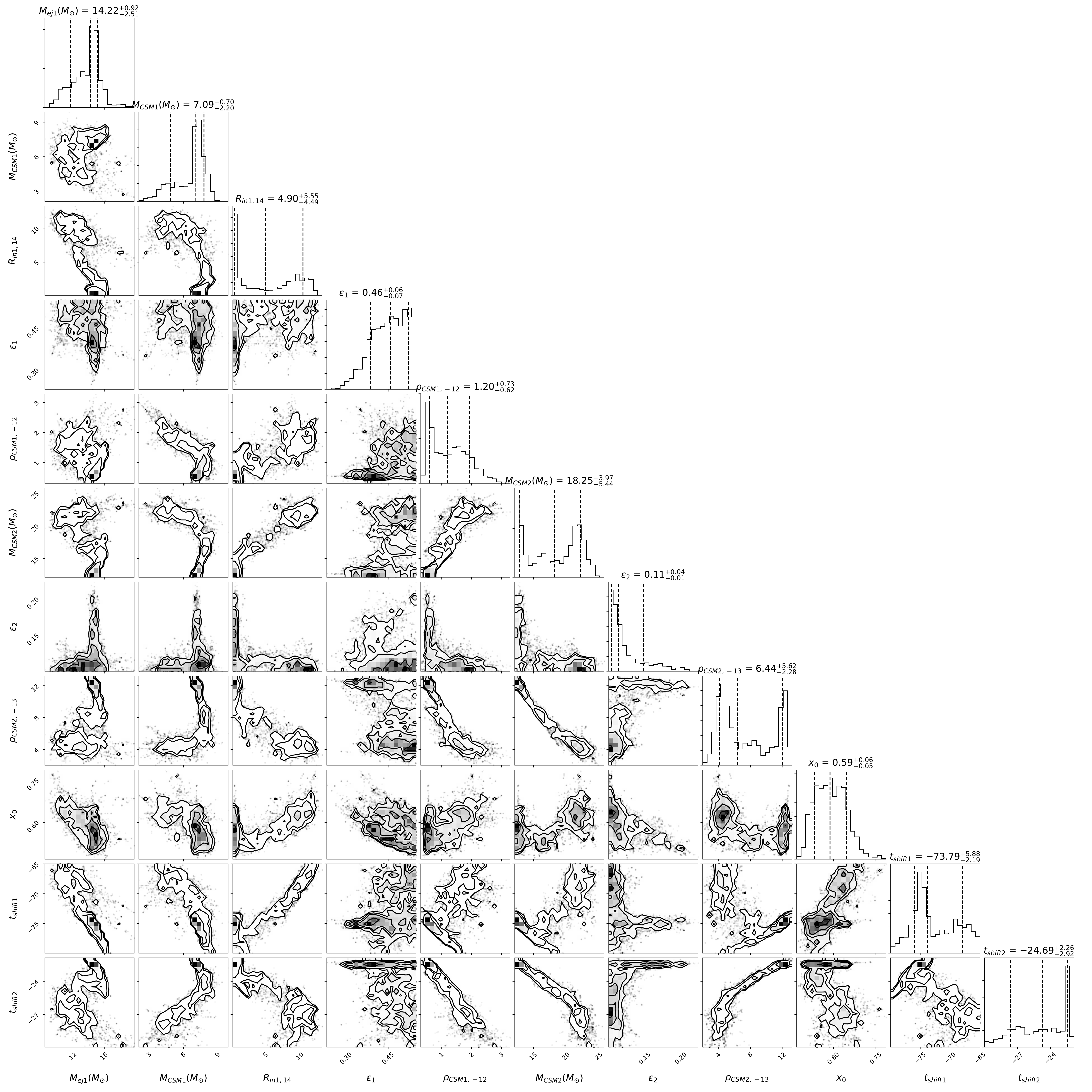}
\end{center}
\caption{Corner plot of the parameters for fitting the LC of iPTF13dcc.
Medians and $1 \sigma$ ranges are labeled.}
\label{fig:corner-13dcc}
\end{figure}

The theoretical LC of iPTF13dcc is shown in Figure \ref{fig:LC_13dcc}. 
The best-fitting parameters and the corresponding confidence
contour corners of iPTF13dcc are shown in Table \ref{tbl:fitting par}
and Figure \ref{fig:corner-13dcc}, respectively. 
The multiple-collision model can match the unusual light curve
of iPTF13dcc ($\chi^{2}$/dof = 2.25). 
The ejecta mass is $M_{\text{ej}} = 14.2M_{\odot}$, and the
masses of CSM shells associated with the first and second collisions are $7.1 M_{\odot}$
and $18.3 M_{\odot}$, respectively. 
Adopting $t_{\text{shell}}\approx R_{\text{in}}/v_{\text{shell}}$
and $v_{\text{shell}} \sim$ 100 km s$^{-1}$, we can infer
that the two shells had been expelled at 1.55 and 21.72
years before the explosion, respectively.

Due to lacking the data before the maximum brightness of the
early bump of iPTF13dcc, it is difficult to determine the rise time of
the first bump. Based on Table \ref{tbl:der} and Figure \ref%
{fig:LC_13dcc}, we find that the rise time of the early bump depends on
the forward-shock termination timescale of the first interaction $t_{\text{%
FS,BO}}=17.8$ days while the decline rate of the late-time LC is
determined by the reverse shock of the second collision.

\subsection{The factors influencing the LC features}

The observed properties of a core collapse SN are determined by
several physical parameters, including the mass of ejecta $M_{\text{ej}}$,
the kinetic energy of the SN ejecta $E_{\text{K}}$, the composition of the
ejecta, and the structure of the envelope of the progenitor at the time of
explosion. These properties result in different types of observed SNe. The
LCs of some multi-peaked SNe show evidence of multiple interactions
between the SN ejecta and the pre-existing CSM shells. In the interaction
model, several parameters related to the CSM, e.g., the mass of CSM
$M_{\text{CSM}}$, the density profile of the CSM, CSM composition,
must be taken into account. Different parameters would lead to different
observational features.

The CSM properties are directly reflected by the LC shape. As shown
in Figures \ref{fig:LC_15esb} and \ref{fig:LC_13dcc}, the
luminosities provided by the forward shocks are usually larger than that
provided by the reverse shocks. The peak times of bumps depend on the
timescales of the forward shocks $t_{\text{FS,BO}}$, while the reverse shock
affects the final decline rate. Less massive CSM tends to power narrower
LC and the larger CSM density results in a slower decline. The peak
luminosity is sensitive to the ejecta mass $M_{\text{ej}}$, so the explosion
of a more massive star yields a brighter peak.

\section{Discussion and Conclusions}

\label{sec:dis}

Massive stars could be unstable and experience mass losses in the form of
eruptions in the final stage of their lives (see,
\citealt{Smi2014} and the references therein). \citet{ofek2014}
pointed out that more than 50\% of the progenitors of type IIn SNe have
experienced at least one pre-explosion eruption. In several cases, the
progenitors of SNe could expel at least two shells and/or winds. Thus, it is
expected that the interactions between the ejecta and multiple shells/winds
would power a bumpy LCs showing two or more peaks.

In this paper, we have studied two such bumpy SNe, iPTF15esb and iPTF13dcc, which
show the undulation features that clearly deviate from the smooth rising and
fading. the LC of iPTF15esb has two peaks and a post-peak plateau, while the
LC of iPTF13dcc shows an early-time bump and a late-time rebrightening. All
the previous energy-source models cannot account for these exotic features.
We suggested that the LC undulations of iPTF15esb could arise from SN ejecta
interacting with multiple dense CSM shells, which may be expelled by the eruptions of
the progenitors. The interaction model for the LC of iPTF15esb is also favored by the broad H$%
\alpha$ emission lines in the late-time spectra which might be produced by
the interaction of SN ejecta with hydrogen-rich CSM shell located at a large
distance from the progenitor star and was ejected by the progenitor star
about 16.4 years before explosion.

To solve these problems, we generalize the ``single" ejecta-CSM interaction model to
the multiple interaction model involving multiple CSM shells and/or winds. 
By employing this new model to fit the LC of iPTF15esb, we got rather good
results and found that the first peak of the LC of iPTF15esb
might be powered by the interaction between the SN ejecta and stellar wind while
both the second peak and the plateau might be powered by the two CSM shells at
different radii. By fitting the LC, we found that the masses
expelled by the progenitor of iPTF15esb are 0.49 $M_{\odot}$ and 1.46$M_{\odot}$
and that the mass-loss rate ($\dot{M}$) of the wind is $0.19-1.9 M_{\odot}$ yr$^{-1}$,
which is comparable to that of SN~1994W ($\dot{M} \sim 0.2 M_{\odot}$ yr$^{-1}$,
\citealt{chu2004}), SN~1995G ($\dot{M} \sim 0.1 M_{\odot}$ yr$^{-1}$, \citealt{chu2003}),
and iPTF13z ($\dot{M} \sim 0.1-2 M_{\odot}$ yr$^{-1}$, \citealt{Nyh2017}).

Furthermore, we also fitted the double-peaked LC of iPTF13dcc using this model
and got rather satisfactory results. In this fit, the LC of iPTF13dcc was
powered by the interactions between the ejecta and two CSM shells whose masses
are 18.3 $M_{\odot}$ and 7.1$M_{\odot}$, respectively. This positive result
suggests that this model is also promising to account for the LCs of several
double-peak SLSNe/SNe (e.g., \citealt{Nic2016,Roy2016,Vre2017}).

It is necessary to discuss the origin of the shells expelled by the
progenitors of iPTF15esb and iPTF13dcc. Several models have been proposed to
explain the violent pre-supernova eruptions. \citet{Woo2007} and \citet%
{Woo2017} suggested that a very massive progenitor may undergo several
episodes of pulsational pair instabilities and eject several massive shells before
the SN explosion. The second possible origin is related to the binary
interaction in which model the shells ejected by the progenitor are supposed
to be formed by the large mass ejections from the progenitor interacting
with its companion star \citep{Pod1992}.
In these successive collisions, the masses of the CSM shells
decrease from the outermost shell (the first eruption) to innermost shell
(the final eruption) but their density increases (since the densities of the
interior of the progenitors are larger than that of the exterior).
This property is consistent with the variation trend of our inferred densities of
the ejected shells and wind of the progenitors of iPTF15esb and iPTF13dcc.

The mass loss of the progenitor of an SN is an important process of stellar evolution.
However, our understanding of the stellar mass loss mechanism remains incomplete.
Further investigations for SLSNe/SNe like iPTF15esb and iPTF13dcc should shed
light on the nature of the mass-loss history of their progenitors.

\acknowledgments We would like to thank an anonymous referee for
constructive suggestions that have allowed us
to improve our manuscript significantly. We also thank Hai Yu,
Weikang Zheng, Bing Zhang, Can-Min Deng, and Xue-Feng Wu for
helpful discussions. This work was supported by the National Basic Research
Program (``973" Program) of China (grant no. 2014CB845800), the National Key R\&D
Program of China (grant no. 2017YFA0402600) and the National Natural Science Foundation
of China (grant no. 11573014). L.J.W. was also
supported by the National Program on Key Research and Development Project of
China (grant no. 2016YFA0400801). L.D.L. and S.Q.W. are supported by China
Scholarship Program to conduct research at UNLV and UCB, respectively.

\clearpage

\begin{sidewaystable}[tbph]
\caption{The fitting parameters for iPTF15esb and iPTF13dcc}
\begin{center}

\begin{tabular}{cccccccccc}
\hline\hline
& $i$th & $s$ & $\kappa $ & $M_{\text{ej}}$ & $M_{\text{CSM}}$ &
$\rho _{\text{CSM,in}}$ \tablenotemark{b} & $\epsilon $ \tablenotemark{c} & $%
t_{\text{tr}}$ & $R_{\text{in}}$ \\
& interaction  &  & $\left( \text{cm}^{2}\text{ g}^{-1}\right) $ & $\left( M_{\odot
}\right) $ & $\left( M_{\odot }\right) $ & $\left( 10^{-13}\text{ g cm}%
^{-3}\right) $ &  & $\left( \text{days}\right) $ & $\left( 10^{15}\text{cm}%
\right) $ \\ \hline\hline
iPTF15esb &  &  &  &  &  &  &  &  &  \\ \hline
& 1 & 2 & 0.2 & 3.92$^{+0.24}_{-0.52}$ & 0.49$^{+0.05}_{-0.03}$ & 20.1$^{+1.9}_{-5.1}$ & 0.42$^{+0.03}_{-0.02}$ & $-8.5$ & 0.22$^{+0.04}_{-0.02}$ \\
& 2 & 0 & 0.2 & 4.41 \tablenotemark{a} & 1.46$^{+0.24}_{-0.24}$  & 4.35$^{+0.45}_{-0.16}$  & 0.19$^{+0.02}_{-0.02}$  & 5.2 & 2.11 \tablenotemark{d} \\
& 3 & 0 & 0.33 & 5.87 \tablenotemark{a} & 2.14$^{+0.13}_{-0.05}$  & 0.43$^{+0.05}_{-0.02}$  & 0.12$^{+0.01}_{-0.01}$  & 24.2 & 5.15\tablenotemark{d} \\ \hline\hline
iPTF13dcc &  &  &  &  &  &  &  &  &  \\ \hline
& 1 & 0 & 0.2 & 14.22$^{+0.92}_{-2.51}$  & 7.09$^{+0.70}_{-2.20}$  & 12.0$^{+7.3}_{-6.2}$  & 0.46$^{+0.06}_{-0.07}$  & -73.79 & 0.1$^{+0.17}_{-0.05}$  \\
& 2 & 0 & 0.2 & 21.31 \tablenotemark{a} & 18.25$^{+3.97}_{-5.44}$  & 6.4$^{+5.62}_{-2.28}$  & 0.11$^{+0.04}_{-0.01}$  & $-24.69$ & 6.85\tablenotemark{d} \\ \hline
\end{tabular}%

\end{center}
\par
a. $M_{\text{ej,2}}$ and $M_{\text{ej,3}}$ are not fitting parameters, but
calculated by Equation (\ref{eq: Mej}).
\par
b. $\rho _{\text{CSM,in}}$ is the density of the CSM at radius $R=R_{\text{in%
}}$.
\par
c. $\epsilon $ is the conversion efficiency from the kinetic energy to
radiation.
\par
d. $R_{\text{in,2}}$ and $R_{\text{in,3}}$ are not fitting parameters, but
calculated by Equation (\ref{eq: Rin}).
\label{tbl:fitting par}
\end{sidewaystable}

\clearpage
\begin{sidewaystable}[tbph]
\caption{The derived physical parameters}
\label{tbl:der}
\begin{center}

\begin{tabular}{cccccccccc}
\hline\hline
& $i$th interaction & $M_{\text{CSM,th}}$\tablenotemark{a} & $t_{\text{FS,BO}%
}$ & $t_{\text{RS,*}}$ & $t_{\text{diff}}$ & $\tau _{\text{CSM}}$ %
\tablenotemark{b} & $R_{\text{out}}$ & $t_{\text{erupt}}$ \tablenotemark{c} & $\dot{M}$
\\
&  & $\left( M_{\odot }\right) $ & $\left( \text{days}\right) $ & $\left(
\text{days}\right) $ & $\left( \text{days}\right) $ &  & $\left( 10^{15}%
\text{ cm}\right) $ & $\left( \text{yr}\right) $ & $\left(M_{\odot} \text{yr}^{-1}\right) $\\ \hline\hline
iPTF15esb &  &  &  &  &  &  &  & & \\ \hline
& 1 & 0.47 & 9.53& 178.7 & 10.0 & 78.8 & 0.97 & - & 0.19$-$1.9\\
& 2 &1.35& 16.8 & 39.7 & 9.0 & 10.1 & 2.25 & 6.69 &  - \tablenotemark{d}
\\
& 3 & 1.77 & 39.6 & 120.0 & 6.2 & 4.1 & 5.5 & 16.34 & - \tablenotemark{d}
\\ \hline\hline
iPTF13dcc &  &  &  &  &  &  &  &  &\\ \hline
& 1 & 7.04  & 12.7& 29.6& 38.9 & 225.5& 1.42  & 1.55 & - \tablenotemark{d}
\\
& 2 & 17.23 & 25.5 & 44.67 & 27.6& 12.14& 6.94 & 21.72& - \tablenotemark{d}
\\ \hline
\end{tabular}%

\end{center}
\par
a. $M_{\text{CSM,th}}$ is the mass of optically thick CSM.
\par
b. $\tau _{\text{CSM}}$ is the optical depth of CSM.
\par
c. $t_{\text{erupt}}$ is the time of the progenitor star erupting the CSM
shells before explosion. Here we assume the velocity of the progenitor wind $%
v_{\text{w}} = 100-1,000$ km s$^{-1}$ and the shell expansion velocities $v_{\text{%
shell}}=100$ km s$^{-1}$.
\par
d. The shells were promptly expelled by some instability process and
their mass-loss rates cannot be calculated.
\end{sidewaystable}

\clearpage

\clearpage

\clearpage

\clearpage

\end{document}